\documentclass[aps,prl,reprint,showpacs,superscriptaddress]{revtex4-1}
\usepackage{amsmath}
\usepackage{textcomp}
\usepackage{graphicx,color}
\usepackage{threeparttable}

\begin{document}

\title{
Interplay of the inverse proximity effect and magnetic field in out-of-equilibrium single-electron devices
}

\author{Shuji Nakamura}
\email[]{shuji.nakamura@aist.go.jp}
\affiliation{National Institute of Advanced Industrial Science and Technology, 1-1-1 Umezono, Tsukuba, Ibaraki 305-8563, Japan}

\author{Yuri A. Pashkin}
\affiliation{Department of Physics, Lancaster University, Lancaster LA1 4YB, UK}

\author{Mathieu Taupin}
\affiliation{Institute of Solid State Physics, Vienna University of Technology, Wiedner Hauptstrasse 8-10, 1040 Vienna, Austria}

\author{Ville F. Maisi}
\affiliation{Low Temperature Laboratory, Department of Applied Physics, Aalto University, 00076 Aalto, Finland}
\affiliation{Center for Quantum Devices, Niels Bohr Institute, University of Copenhagen, Universitetsparken 5, 2100 Copenhagen {\O}, Denmark}

\author{Ivan M. Khaymovich}
\email[]{ivan.khaymovich@pks.mpg.de}
\affiliation{Max Planck Institute for the Physics of Complex Systems, D-01187 Dresden, Germany}
\affiliation{Institute for Physics of Microstructures, Russian Academy of Sciences, 603950 Nizhny Novgorod, GSP-105, Russia}

\author{Alexander\ S.\ Mel'nikov}
\affiliation{Institute for Physics of Microstructures, Russian Academy of Sciences, 603950 Nizhny Novgorod, GSP-105, Russia}
\affiliation{Lobachevsky State University of Nizhny Novgorod, 23 Prospekt Gagarina, 603950, Nizhny Novgorod, Russia}

\author{Joonas T. Peltonen}
\affiliation{Low Temperature Laboratory, Department of Applied Physics, Aalto University, 00076 Aalto, Finland}

\author{Jukka P. Pekola }
\affiliation{Low Temperature Laboratory, Department of Applied Physics, Aalto University, 00076 Aalto, Finland}

\author{Yuma Okazaki}
\affiliation{National Institute of Advanced Industrial Science and Technology, 1-1-1 Umezono, Tsukuba, Ibaraki 305-8563, Japan}

\author{Satoshi Kashiwaya}
\affiliation{National Institute of Advanced Industrial Science and Technology, 1-1-1 Umezono, Tsukuba, Ibaraki 305-8563, Japan}

\author{Shiro Kawabata}
\affiliation{National Institute of Advanced Industrial Science and Technology, 1-1-1 Umezono, Tsukuba, Ibaraki 305-8563, Japan}

\author{Andrey S. Vasenko}
\affiliation{National Research University Higher School of Economics, 101000 Moscow, Russia}

\author{Jaw-Shen Tsai}
\affiliation{RIKEN Center for Emergent Matter Science, Wako, Saitama 351-0198, Japan
}
\affiliation{Department of Physics, Tokyo University of Science, Kagurazaka, Tokyo 162-8601, Japan}

\author{Nobu-Hisa Kaneko}
\affiliation{National Institute of Advanced Industrial Science and Technology, 1-1-1 Umezono, Tsukuba, Ibaraki 305-8563, Japan}

\date{\today}

\begin{abstract}

The magnetic field is shown to affect significantly non-equilibrium quasiparticle (QP) distributions under conditions of inverse proximity effect on the remarkable example of a single-electron hybrid turnstile.
This effect suppresses the gap in the superconducting leads in the vicinity of turnstile junctions with a Coulomb blockaded island, thus, trapping hot QPs in this region.
Applied magnetic field creates additional QP traps in the form of vortices or regions with reduced superconducting gap in the leads resulting in release of QPs away from junctions.
We present clear experimental evidence of such interplay of the inverse proximity effect and a magnetic field revealing itself in the superconducting gap enhancement in a magnetic field as well as in significant improvement of the turnstile characteristics.
The observed interplay of the inverse proximity effect and external magnetic field, and its theoretical explanation in the context of QP overheating are important for various superconducting and hybrid nanoelectronic devices, which find applications in
quantum computation, photon detection and quantum metrology.
\end{abstract}


\maketitle

\begin{figure}[h!]
	\begin{center}
		\includegraphics[clip,width=\columnwidth]{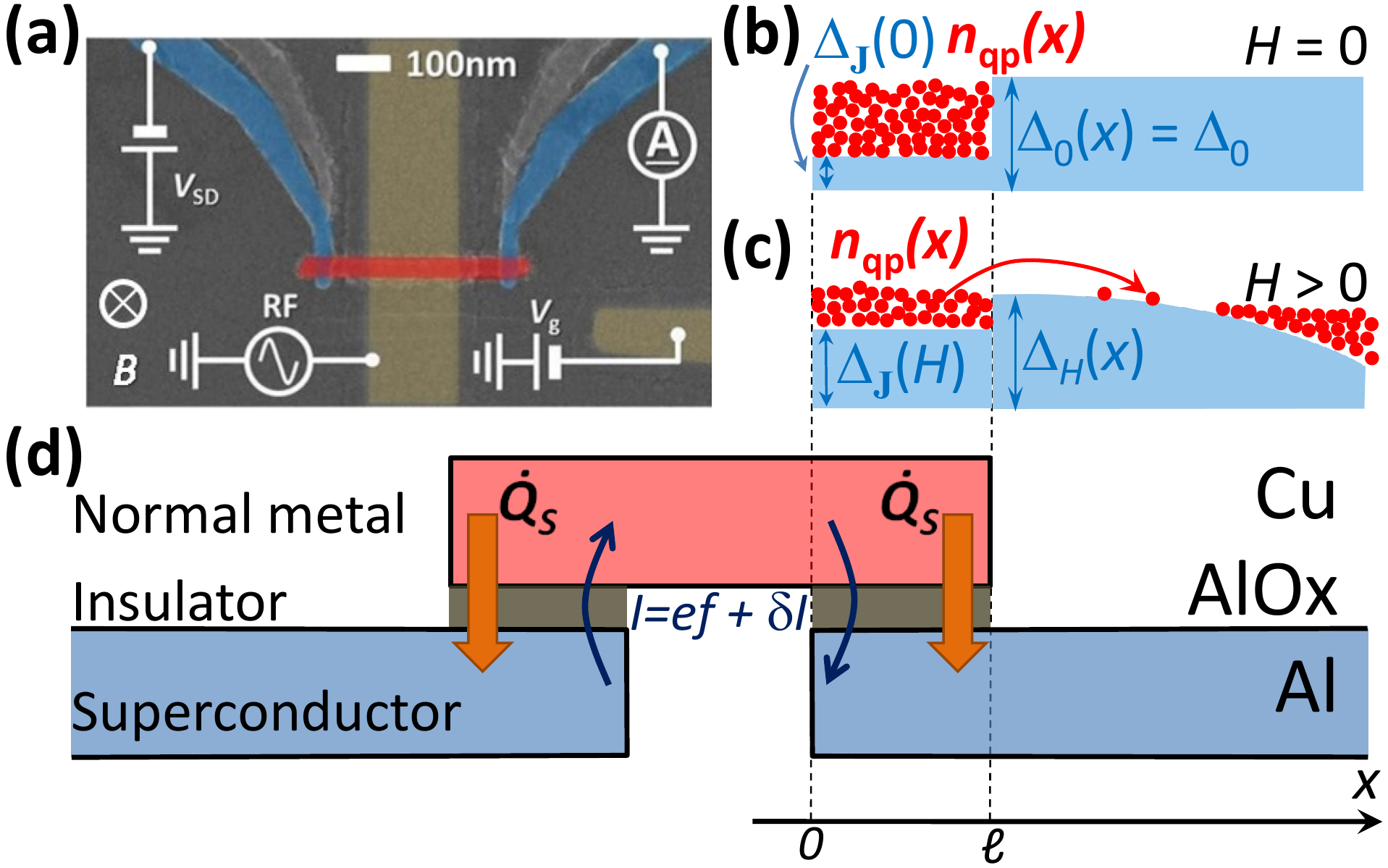}
		\caption{\label{fig:Layout+cartoon} 
(a)~SEM image of the SINIS turnstile and measurement setup.
False color identifies superconducting Al leads (blue), the normal metal Cu island (red),
and the DC side gate and RF bottom gate electrodes (both yellow).
(b),(c) The gap profile $\Delta_{H}(x)$ and $\Delta_{\rm J}=\Delta_H(0)$ (blue shading) and the QP distribution $n_{\rm qp}(x)$ (red circles) in the leads at different magnetic field $H$ values.
(d) Schematic cross-section of the structure with charge (blue arrows) and heat (orange arrows) currents.
In panels (b)~-~(d) the overlap junctions are located in the interval $0<x<\ell$ along $x$-direction.}
	\end{center}
\end{figure}

Proximity effect which induces superconducting correlations into a normal-metal conductor at the interface between a superconductor and a normal-metal conductor has been widely explored and  plays an important role in the physics of superconductors \cite{deGennes_book}.
Considering recent developments one can see that this phenomenon provides a basis for the engineering and manipulation of the
symmetry of the induced superconducting pairing in various hybrid structures including topologically nontrivial systems known to host Majorana fermions \cite{V.MourikScience2012} and possess other exciting properties \cite{EschrigNaturePhys2008,GiazottoNaturePhys2010,Ronzani2014}.
The proximity phenomenon has an important counterpart, namely the inverse proximity effect, which is responsible for the reduction of the superconducting order parameter due to the penetration of electrons through the superconductor~-~normal~metal interface \cite{Gennes1964}.
Microscopically both the proximity and inverse proximity effects can be understood
in terms of the Andreev reflection at the interface of a superconductor and a normal metal \cite{Tinkham_book}.

Reducing the superconducting gap the inverse proximity effect provides a physical mechanism responsible for the
formation of the traps for the nonequilibrium quasiparticles (QPs) known to affect the performance of many
superconducting devices such as X-ray detectors~\cite{Golubov1993,Golubov1994},
single photon detectors~\cite{Goltsman2001},
refrigerators based on normal metal (N)~-~insulator (I)~-~superconductor (S) junctions~\cite{Peltonen2011},
superconducting resonators~\cite{Nsanzineza2014},
superconducting qubits~\cite{Wang2014}, and single-electronic hybrid turnstiles \cite{Taupin2016}.
The problem of heat evacuation in these devices is usually solved by introducing
QP traps based on normal metal inclusions \cite{Goldie1990,Ullom2000,Rajauria2012}, on the local order parameter suppression by an external magnetic field \cite{Peltonen2011,Nsanzineza2014,Wang2014,Taupin2016}
or by using an alternative device design immune to QP overheating \cite{vanZanten2016}.
The traps arising from the inverse proximity phenomenon can be effective, of course, only for the relatively low resistive devices
since the noticeable gap reduction at the SN interface requires quite transparent interfaces.
On one hand, due to location in the working region of the device such traps confine QPs very effectively and may strongly spoil the device operation.
On the other hand, the inhomogeneous spatial gap profile in the leads can be utilized in such devices as X-ray detectors~\cite{Golubov1993,Golubov1994} where gap engineering is needed to increase the signal from the QPs located at the junction.

In this Letter we address an intriguing possibility to diminish the effect of the QP traps formed due to the inverse proximity phenomenon using the competition in QP confinement between these traps and the ones formed in the presence of an applied magnetic field.
This interplay between different trapping mechanisms and resulting redistribution of the nonequilibrium QPs
is a rather general phenomenon interesting both as a manifestation of fundamental properties of superconducting correlations and as
a trick useful for applications. The interplay mentioned above reveals itself in a nontrivial counter-intuitive
consequence: the superconducting gap at the interfaces appears to increase
with the increase in magnetic field.
Both the inhomogeneous gap profile and its sensitivity to magnetic field can have applications to the superconducting or hybrid devices
where several superconductors with different gap values are usually used \cite{Golubov1993,Golubov1994}.
The main advantages of the considered gap engineering effect comparing to the usual solution are
the perfect matching of the superconducting parts with different gaps without barriers or interface potentials and
the tunability of the gap profile through the applied magnetic field.

To observe the interplay between the magnetic field and inverse proximity effect experimentally we
study the charge transport through a hybrid turnstile~\cite{Pekola2007}, i.e.,
a single-electron transistor consisting of a small normal metal island
in Coulomb blockade regime with the charging energy $E_{C}$
 tunnel-coupled to voltage biased superconducting electrodes (SINIS)
and controlled by both DC and RF gate voltages (see Fig.~\ref{fig:Layout+cartoon}(a)).
Comparing the samples with different tunnel resistances $R_{\rm T}$ we clearly demonstrate
that the decrease of $R_{\rm T}$ down to the order of the resistance quantum $R_{\rm Q} = \frac{h}{e^{2}} \sim 25.8$~k$\Omega$,
i.e. the increase of the junction transparency enhances the excess current $\delta I = I - e f$ in the turnstile regime,
dominated by the hot QP contribution. Here the product $e f$ of the elementary electron charge $e$ and the drive frequency $f$ is the ideal value
of the current $I$ in the turnstile regime. The decrease of $R_{\rm T}$ also affects the superconducting gap $\Delta_{\rm J}$
at the junctions, $0<x<\ell$, keeping the gap $\Delta_0(x)=\Delta_0$ away from the junctions intact, see Fig.~\ref{fig:Layout+cartoon}(b).
$x$ is a coordinate along the leads, $\ell$ is the size of the proximitized region at the SIN junctions (Fig.~\ref{fig:Layout+cartoon}(d)).
A weak magnetic field perpendicular to the sample recovers $\Delta_{\rm J}(H)$ closer to its bulk value $\Delta_0$
and reduces simultaneously the turnstile excess current.
This is consistent with the developed theoretical picture of simultaneous
weakening of the inverse proximity effect and
release of hot QPs from the vicinity of the junctions
due to the reduction of the gap $\Delta_{H}(x)$ away from the junctions,
mediated by magnetic field (see Fig.~\ref{fig:Layout+cartoon}(c)).
This theoretical model
explains the magnetic-field dependence of the excess current
and gives semi-quantitative agreement with the experimental data.

The SINIS samples are fabricated with the standard electron-beam lithography and shadow evaporation technique~\cite{Fulton_Dolan1987},
see scanning electron microscope (SEM) micrograph of it in Fig.~\ref{fig:Layout+cartoon}(a)).
The Cu island ($30$~nm thick) is connected to two Al leads ($18$~nm thick) via a thin insulating aluminum oxide layer
formed by exposing the device to oxygen atmosphere just after the deposition of Al.
The transparency of AlO$_x$ barrier is controlled by varying the oxygen pressure and exposing time (see~\cite{SM} for fabrication details).
We fabricate and measure three devices with different tunnel resistances as listed in Table~\ref{tab:example}.

\begin{figure}[t]
	\begin{center}
		\includegraphics[clip,width=\columnwidth]{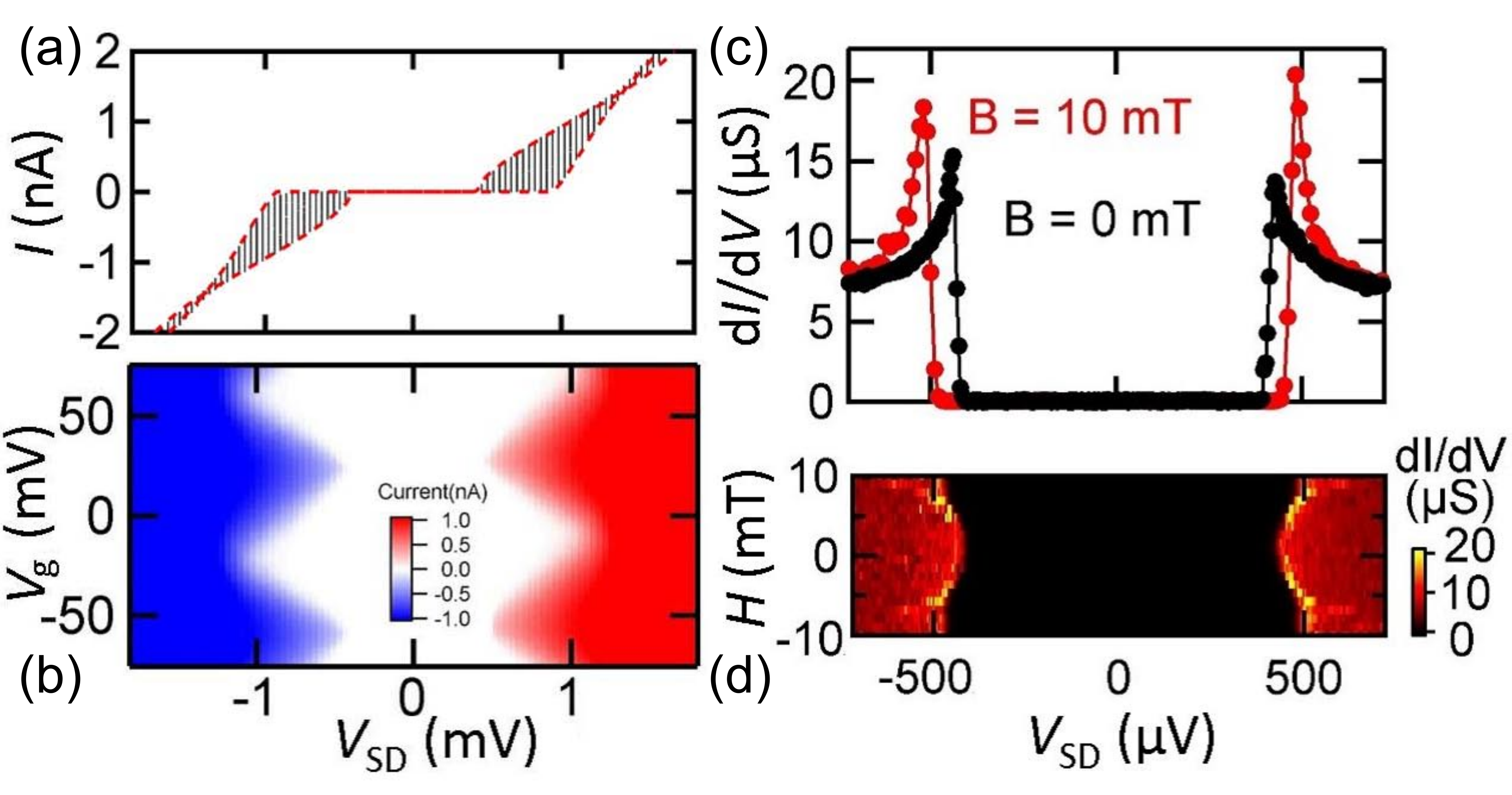}
		\caption{\label{fig:IVs}
(a) $I-V$ characteristics of sample H for various dc gate voltages $V_{\rm g}$.
Dashed red lines are simulated $I-V$ characteristics for the gate-open and closed states;
(b) Source-drain turnstile current vs $V_{\rm g}$ and $V_{\rm SD}$ forming Coulomb blockade diamonds;
(c) Differential conductance of the sample R as a function of bias voltage at zero (black) and finite (red) magnetic fields $H$;
(d) Color plot of differential conductance vs $V_{\rm SD}$ and $H$.
}
	\end{center}
\end{figure}

All measurements are performed in dilution refrigerators at a base temperature $T_0$ around $100$~mK in the
the measurement setup, schematics of which is shown in Fig.~\ref{fig:Layout+cartoon}(a).
An applied DC gate voltage ($V_{\rm g}$) tunes offset charges on the normal metal island, and an additional sinusoidal signal with the amplitude $A_{\rm g}$ is applied to the RF gate for the electronic pumping.
The pumping experiments are carried at a drive frequency $f=100$~MHz with
the DC gate voltage $V_{\rm g}$ fixed at a gate-open state (offset charge of the turnstile island is 0.5) and
the source-drain voltage tuned to the optimal point $eV_{\rm SD} = \Delta$ for turnstile operation~\cite{Pekola2007}.
The current is measured with a room-temperature current amplifier calibrated by a standard resistor and a calibrated voltage source.

\begin{figure*}[t]
	\begin{center}
		\includegraphics[clip,width=0.8\textwidth]{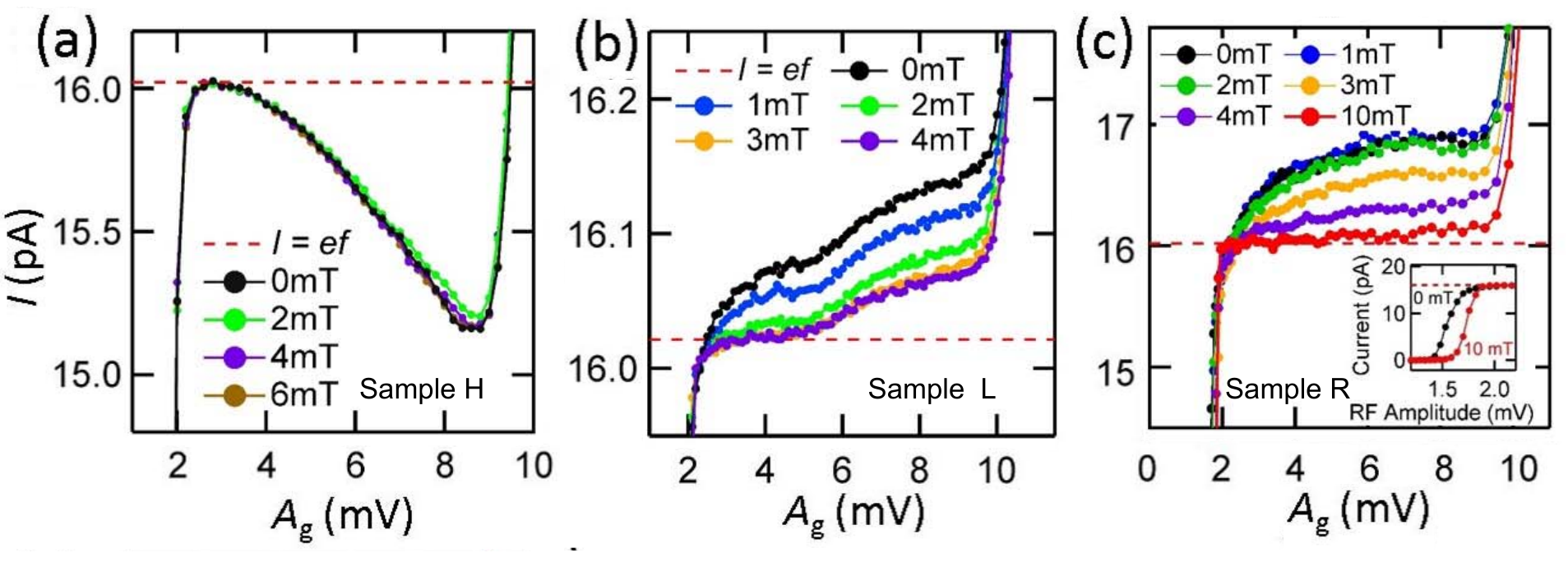}
		\caption{\label{fig:I_vs_Ag}
Pumped current in turnstile regime (colored dots) of (a) sample H, (b) sample L, and (c) sample R vs
RF gate amplitude at different magnetic fields.
Horizontal red dashed lines show the ideal value of the pumped current $I = e f$ at the turnstile frequency $f=100$~MHz.
(panel c inset) a close-up of the onset of the current plateau at 0~mT and 10~mT.
			}
	\end{center}
\end{figure*}

The sample parameter values $R_{\rm T}$, $E_C$, and $\Delta_0$
have been extracted from the $I-V$ characteristics using standard numerical simulation based on the Fermi's golden rule and the master equation~\cite{KemppinenEPJ2009}, see typical $I-V$ curves at different $V_{\rm g}$ for sample H in Fig.~\ref{fig:IVs}(a), (b).
The parameter values for all samples are summarized in Table~\ref{tab:example}.
\begin{table}[t]
	\caption{\label{tab:example}Parameters of the fabricated SINIS turnstiles.}
	\begin{ruledtabular}
		\begin{threeparttable}
			\begin{tabular}{cccc}
			Sample    & $R_{T}$, k$\Omega$  & $E_{C}/\Delta_{0}$  & $\Delta_{0}$, $\mu$eV \\ \hline
			H         & 230      & 1.6   & 216\\
			L         & 55       & 1.6   & 215\\
			R         & 60       & 1.8   & 210\\
			\end{tabular}
				\begin{tablenotes}\footnotesize
					\item Samples H and L are fabricated and measured in setup~1, while sample~R is fabricated and measured in setup~2.
(See Supplemental Material for details \cite{SM}.)
			    \end{tablenotes}\footnotesize
		\end{threeparttable}	
	\end{ruledtabular}
\end{table}

Starting with a zero magnetic field we observe that
the turnstile current for highly resistive sample H with $R_{\rm T}\sim 9 R_{\rm Q}$ (black dots in Fig.~\ref{fig:I_vs_Ag}(a)) demonstrates a negative deviation from its ideal value $I = e f$ (red dashed line) at high $A_{\rm g}$ values in the full agreement with the effect of backtunneling processes at high resistance/Coulomb energy~\cite{Kemppinen2009}.
The current $I$ in low resistive samples L and R (black dots in panels (b) and (c)) with $R_{\rm T}\sim 2R_{\rm Q}$ exceeds $e f$ by 0.7~\% and 5~\%, respectively.

As expected, a weak magnetic field applied perpendicular to the substrate has almost no effect on the turnstile current in sample H (color dots in Fig.~\ref{fig:I_vs_Ag}(a)), whereas it reduces significantly the excess current in low resistive samples (color dots in panels (b)--(c)).
Simultaneously with the reduction of the excess current in samples L and R, the RF threshold amplitude value $A_{\rm g}^{\rm th}=\Delta_{\rm J}/e - V_{\rm SD}/2$ increases (see inset of Fig.~\ref{fig:I_vs_Ag}(c)), although the dc source-drain and gate voltages are kept unchanged.
This points out that superconducting gap value $\Delta_{\rm J}$ at the junctions is enhanced by an applied magnetic field.
This increment of $\Delta_{\rm J}(H)$ in the field is observed explicitly by measuring the differential conductance $dI/dV$ in the gate-open state, see plots for sample R in Fig.~\ref{fig:IVs}(c), (d).
The gap increases by more than $10$~\% at $H=10$~mT relatively to its zero field value.
The enhancement of the superconducting gap in sample L is qualitatively the same (not shown).

Our quantitative theoretical description is based on the idea schematically shown in Fig.~\ref{fig:Layout+cartoon}(b), (c).
At zero field hot QPs produced due to turnstile operation are trapped at the junctions in region with reduced gap $\Delta_{\rm J}$ (panel (b))
and cause the excess current~\cite{Knowles2012}.
Due to the lead geometry magnetic field $H$ reduces the gap in the wide region away from the junctions
(or even suppresses it due to vortex penetration),
opens the way for QPs to escape from the trap and therefore diminishes the overheating of the proximitized region (panel (c)).
Reduction of both overheating and the inverse proximity effect enhances the gap $\Delta_{\rm J}$.
Further increase of the field leads to the negative effect on $\Delta_{\rm J}$ diminishing therefore the turnstile accuracy.
In the theoretical model we take the following assumptions:
(i)~The electron-electron relaxation rate is larger than the drive frequency and tunneling rates keeping electronic distributions in the leads to be of the Fermi-Dirac form with effective spatially dependent temperature $T(x)$ \cite{Footnote_QEq}
;
(ii)~The excess current $\delta I = I - e f$ is dominated by the overheating contribution (see, e.g.,~\cite{Saira2012, Maisi2014, Taupin2016})
\begin{equation}
\label{eq:QP_current_general}
{\delta}I = C\frac{\sqrt{2\pi\Delta_{\rm J}k_{\rm B}T_{\rm J}}}{eR_{T}}\exp\left[-\frac{\Delta_{\rm J}}{k_{\rm B}T_{\rm J}}\right] \equiv \frac{C\Delta_{\rm J}}{eR_{T}}N_{\rm qp}(0)
\end{equation}
with a numerical prefactor $C(A_{\rm g})\sim 1$ determined by the wave-form and the amplitude $A_{\rm g}$ of the RF gate voltage;
(iii)~The temperature $T(x)=T_{\rm J}$, the gap $\Delta_{\rm J}(H, x)=\Delta_{\rm J}(H)$, and the normalized QP number $N_{\rm qp}(x)$ in the proximitized region, $0<x<\ell$, are constant~\cite{Footnote_const_Nqp};
(iv)~Most of the dissipated heat $I V\simeq 2\dot{Q}_S$ goes to the leads and keeps N-island close to the equilibrium.

Due to the lead geometry the stationary temperature profile $T(x)$ is determined by the solution of the one-dimensional approximation of the heat diffusion equation
\begin{equation}
\label{eq:heat_diffusion_eq}
\frac{\partial}{\partial x}\left[\kappa_{\rm S}(T(x),x)\frac{\partial}{\partial x}T(x) \right] = \dot{q}_{\rm e-ph}(T(x))
\end{equation}
with the following boundary conditions at the junctions, $0<x<\ell$, $T|_{\rm J}=T_{\rm J}$ and away from it, $x\to\infty$
\begin{subequations}\label{eq:bound_conds}
\begin{align}
\left.-\kappa_{\rm S}(T(x),x)\frac{\partial}{\partial x}T(x)\right|&_{\rm J} = \dot{Q}_{S}/A \label{eq:heat_flow}  \ ,\\
\left.T(x)\right|&_{x \to \infty} = T_{0}  \label{eq:tempcontinuity} \ .
\end{align}
\end{subequations}
Here $T_0$ is the phonon bath temperature, $\kappa_{\rm S}(T(x),x) \sim \frac{2\Delta_{H}^2(x)}{e^2\rho_{n}T}e^{-\Delta_{H}(x)/k_{\rm B}T}$ is the thermal conductivity of the superconductor, $A$ is the junction area, $\rho_{n}=30$~n$\Omega$m is the normal state resistivity of the superconductor~\cite{Knowles2012}, $\dot{q}_{\rm e-ph}$ is the density of the electron-phonon relaxation term approximated as $\dot{q}_{\rm e-ph} \sim \Sigma T^{5} e^{-\Delta_{H}(x)/k_{\rm B}T} $ in the low temperature limit $T_0 \ll T \ll \Delta_{H}/k_{\rm B}$ with $\Sigma \simeq 3 \times 10^8$~W~K$^{-5}$~m$^{-3}$ being the electron phonon material parameter~\cite{Kautz1993, GiazottoRMP2009, Maisi2013, Taupin2016}.
Using continuity conditions for the heat flow and the temperature at the edge of the proximitized region $x = \ell$ analogous to Eqs.~\eqref{eq:bound_conds} one can find the solution for $T_{\rm J}$ and $N_{\rm qp}(\ell)$ as (cf.~\cite{Knowles2012})
\begin{equation}
\label{eq:Numberquasi_without}
N_{\rm qp}(\ell) = \sqrt{\frac{2\pi k_{\rm B}T_{\rm J}}{\Delta_{\ell} }}e^{-\frac{\Delta_{\ell}}{k_{\rm B}T_{\rm J}}}
\simeq \frac{e^2\dot{Q}_{S}\rho_{n}\mathcal{R}_H[w(x)]}{\ell  \sqrt{2 k_{\rm B}T_{\rm J}\Delta_{\ell}^3/\pi}}
\end{equation}
with the superconducting gap just behind the junctions $\Delta_{\ell} = \Delta_{H}(\ell )$ and the resistance $\mathcal{R}_H[w(x)]$ of the lead normalized to the sheet resistance being a functional of the lead width profile $w(x)$~\cite{SM}.
In the limit of small excess current $\delta I \ll e f$ one can neglect its contribution to the heat flux rate $\dot{Q}_{\rm S} \simeq e f V/2$ and
due to the exponential sensitivity of the left-hand side of Eq.~\eqref{eq:Numberquasi_without} to $T_{\rm J}$, one can disregard polynomial $T_{\rm J}$-dependence in the right-hand side.
As a result, the excess current normalized to the ideal current value $e f$ can be analytically estimated as follows
\begin{equation}
\label{eq:QP_current}
\frac{\delta I(H)}{e f} \simeq \frac{C\sqrt{2\pi k_{\rm B}T_{\rm J}\Delta_{\rm J}}}{e^{2}R_{T}f}\left[\frac{eV_{\rm SD}}{4\Delta_{\ell}}\frac{e^{2}\rho_{n}f}{\ell k_{\rm B}T_{\rm J}}\mathcal{R}_H\right]^{\Delta_{\rm J}/\Delta_{\ell}} \ .
\end{equation}
Due to the smallness of the term in square brackets the ratio $\Delta_{\rm J}(H)/\Delta_{\ell}$ plays a significant role in the magnetic-field dependence of $\delta I/(e f)$.
The $H$-dependence of the terms in brackets is taken into account through the geometrical factor $\mathcal{R}_H$ which varies from $\mathcal{R}_0 \simeq 35$ to $\mathcal{R}_H \simeq 20$ at $H = 10$~mT~\cite{SM} due to the vortex penetration into superconducting leads~\cite{Peltonen2011}.

\begin{figure}[b]
	\begin{center}
		\includegraphics[clip,width=\columnwidth]{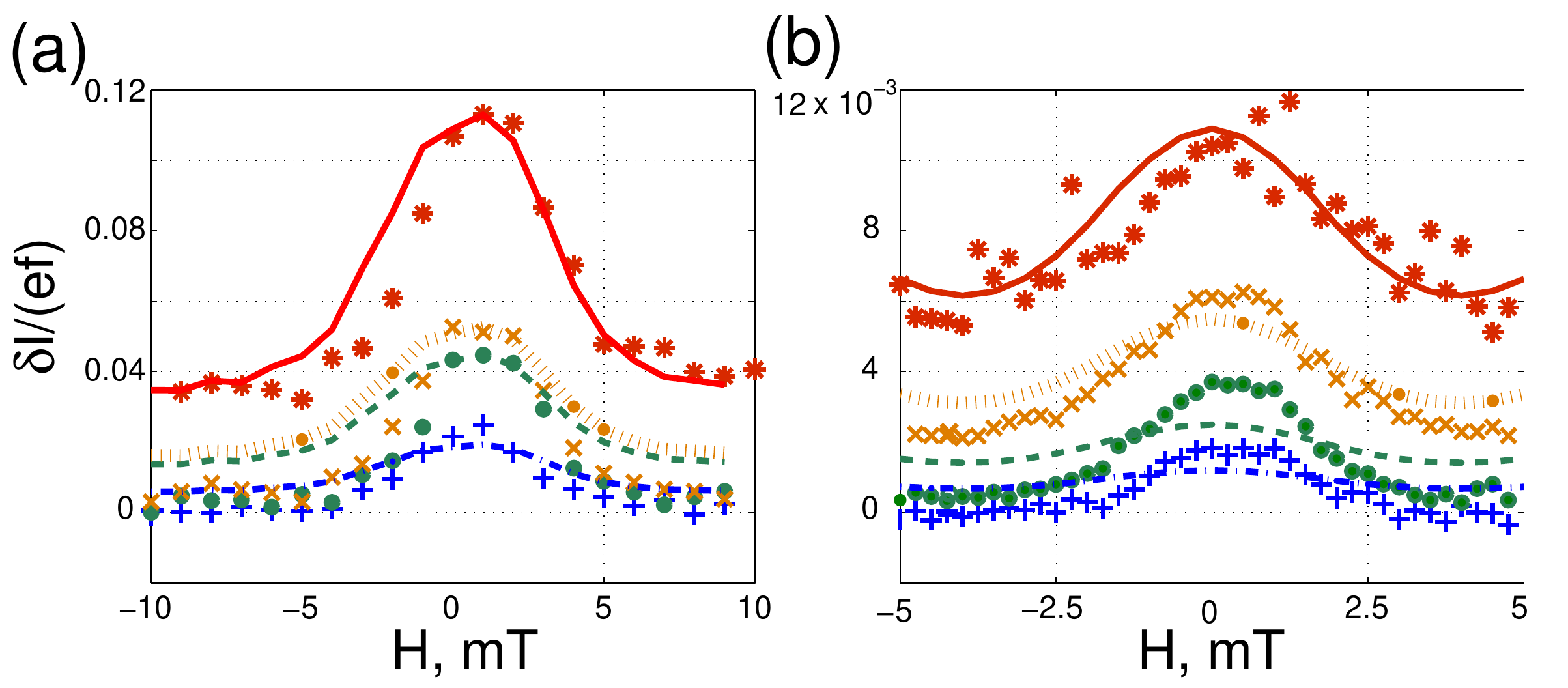}
		\caption{\label{fig:dI_vs_H}
The excess current $\delta I = I - e f$ normalized to $e f$ through (a) sample R and (b) sample L vs magnetic field
at $A_{\rm g} = 3, 5, 7, 10$~mV (from bottom to top).
The experimental data (symbols) is a cross-section of the curves in Fig.~\ref{fig:I_vs_Ag}(b), (c) at fixed $A_{\rm g}$.
Theoretical curves (lines) are obtained using Eqs.~(\ref{eq:QP_current_general}, \ref{eq:Numberquasi_without}).
The zero-field gap $\Delta_{\rm J}(0)/\Delta_0$ at the junctions normalized to its bulk value $\Delta_0$
is fitted as (a) $0.6$ and (b) $0.8$ for $A_{\rm g} = 10$~mV (red), the prefactor is taken to be $C=1$.
Other theoretical curves are fitted with respect to $C(A_{\rm g})$ with the values:
(a) $C=0.17, 0.40, 0.47, 1$, and (b) $C = 0.1, 0.2, 0.5, 1$ (from bottom to top).
}
	\end{center}
\end{figure}

The theoretical values of the excess current shown in Fig.~\ref{fig:dI_vs_H} are calculated by the substitution of the self-consistent solution of~\eqref{eq:Numberquasi_without} for $T_{\rm J}$ into the general formula~\eqref{eq:QP_current_general} with the fitting parameters $\Delta_{\rm J}(0)/\Delta_{\ell}$ common for all $A_{\rm g}$ and the amplitude-dependent factor $C(A_{\rm g})$ \cite{SM}.
The gap profile $\Delta_{\rm J}(H)/\Delta_{\rm J}(0)$ 
is taken from the DC experimental data, see Fig.~\ref{fig:IVs}(c), (d).
Numerical calculation reproduces the experimental results semi-quantitatively both in sample R and sample L.
The data at the largest amplitude $A_g=10$~mV is fitted with $C(A_g)=1$.
The deviation of the current from the theoretical curves at smaller amplitudes and at larger values of magnetic fields
is possibly related to other contributions such as backtunneling, cotunneling, and Andreev processes as the quasiparticle contribution is suppressed.
This effect is more significant in sample L (panel (b)) where the excess current is $\sim 10$ times smaller.
An order of magnitude difference in the amplitude and the different $A_g$-dependence of the excess current in samples R and L with close parameters, see Table~\ref{tab:example},
could be related to the different experimental environment, namely, sample holder shielding and RF wire filtering in 
setups~1 and~2 causing the different profiles of an actual ac gate voltage applied to the sample.

The only subtle point not covered by the developed theoretical model is the $13$~\% increase of the superconducting gap at the junctions $\Delta_{\rm J}(H)/\Delta_{\rm J}(0)$ at weak magnetic fields.
Instead, the theory takes the magnetic field dependence of the gap from the experimental data and shows its evident relation with the excess current.
This gap increase may originate from the enhancement of the order parameter related to the hot quasiparticle density through self-consistency equation.
Another possible reason of the increase of $\Delta_{\rm J}(H)/\Delta_{\rm J}(0)$ is the absence of
the $H$-mediated superfluid velocity in the dead end of the superconducting lead close to the junctions.
Indeed, in the absence of the superfluid velocity the depairing parameter is negligible and both the gap and the order parameter are immune to the field.

To conclude, we study single electron pumping in the SINIS turnstile affected simultaneously by the inverse proximity effect and magnetic field.
In the samples with a low junction resistance a puzzling magnetic-field dependence of the turnstile current and
non-monotonic magnetic-field profile of the superconducting gap at the junctions are observed,
while in the high-resistive sample such effects are absent.
This puzzle is resolved by the theoretical modeling taking into account both
the inverse proximity effect leading to the quasiparticle trapping at the junctions
and the overheating of the superconducting leads by hot quasiparticles resulting in the excess current.
Perpendicular magnetic field releases hot quasiparticles from the proximitized region
by suppressing of the superconductivity away from the junctions and simultaneously weakening the proximity effect.

\begin{acknowledgments}
We acknowledge fruitful discussions with Yoshitake Takane and Srinivas Gandrothula.
This work is partially supported by KAKENHI(16H06090),
by Academy of Finland, 
Project Nos. 284594, 
272218 (M.T., J.T.P., and J.P.P.), 
by the UK EPSRC, Grant No. EP/K01675X/1 (Yu.A.P) and the Royal Society, Grant No. WM110105 (Yu.A.P),
by the Russian Foundation for Basic Research (I.M.K. and A.S.M.), by the Russian Science Foundation, Grant No. 15-12-10020 (I.M.K. and A.S.M.),
and
by the project T3-97 ``Macroscopic Quantum Phenomena at Low Temperatures'',
carried out within the framework of the Basic Research Program at the National Research University Higher School of Economics (HSE) in 2016 (A.S.V.).
\end{acknowledgments}

\section{Supplemental Material}
\subsection{Details of fabrication processing and measurement setup}

The SINIS devices are fabricated on a thermally oxidized SiO$_{2}$ wafer using electron-beam lithography and metal deposition including angle evaporation. Before the final fabrication step in which the SINIS structures were formed, the wafer went through several processing steps to prepare bonding pads and a ground plane.
Each chip contains Ti(5\,nm)/Au(95\,nm) bonding pads and a Ti(5\,nm)/Au(50\,nm) ground plane that has a slot to accommodate an RF line. The RF line was extended to the center of the chip with a 30\,nm-thick and 200\,nm-wide Au strip. The whole wafer was then covered by a layer of SiO$_2$ using spin-on glass which was patterned to open the contact pads. Finally, we fabricate the SINIS turnstiles using a tri-layer resist structure (copolymer resist (400\,nm)/Ge(20\,nm)/PMMA resist (50\,nm)) which is formed by electron beam lithography and dry etching. SINIS devices are connected to the bonding pads by Al leads stretching from the chip center to the bonding pads above the ground plane, thus a large capacitance is formed between the DC leads and the ground plane protecting sensitive SINIS pumps from the electromagnetic noise penetrating into the sample package. Al/AlO$_x$/Cu tunnel junctions of the SINIS pumps are formed by the overlap of 18\,nm-thick Al leads and 30\,nm-thick Cu island deposited in the e-gun evaporator at different angles through a suspended Ge mask formed by electron-beam lithography and dry etching. Aluminum oxide layer on the surface of the deposited Al film is grown by letting pure oxygen or Ar\,+\,O$_2$ mixture into the sample chamber. The tunnel junction resistance is controlled by varying the oxygen/argon pressure and oxidation time.
The normal metal island is capacitively coupled to the RF bottom gate electrode, isolated from the island by the SiO$_{2}$ layer. To optimize the bias voltage for the turnstile operation, we measured the ${I-V}$ characteristic at the gate-open state (offset charge of the island is 0.5) for the estimation of superconducting gap $\Delta$ and applied the bias voltage $V_{\rm SD} = \Delta$.
	
In this experiment, we use two different setups for oxidation and sample characterization (setup 1 and setup 2).
Sample H and sample L are fabricated and measured in setup~1. 
Aluminum was oxidized under static conditions.
(A small amount of gas is introduced into the vacuum chamber).
The oxidation conditions are  40\,s under 37.5\,mTorr of pure oxygen for sample~H, and 2 minutes under 97\,mTorr Ar\,+\,O$_2$(1\,\%) for sample~L.
The measurement is performed in a homemade dilution refrigerator whose base temperature is around 100\,mK. The source-drain voltage ($V_{\rm SD}$) between the superconducting leads is applied using a commercial voltage source (SIM928), and the current is measured with a room-temperature current amplifier (DDPCA 300 from Femto) calibrated by a standard resistor and a calibrated voltage source in the Metrology Institute of Finland (MIKES). The DC signals ($V_{\rm SD}$ and $V_{\rm g}$) are filtered with Thermocoax cables, and the RF line has a $-20$\,dB attenuator at 4.2\,K, thermalized in the helium bath, and a $-20$\,dB attenuator at room temperature.
Sample~R is fabricated and measured in setup~2.
Aluminum was oxidized in a continuous gas flow regime at a constant gas pressure maintained by an automatic pressure regulator.
The oxidation condition is 2 minutes under 30\,mTorr Ar~+~O$_{2}$(10 \%). The measurement is performed in a commercial dilution refrigerator (Oxford Instruments Kelvinox 100) whose base temperature is also about 100\,mK. The voltage source and the current amplifier are same as those in setup~1.
Calibration of the current amplifier is done in the Japanese Metrological Institute (AIST). The DC signals ($V_{\rm SD}$ and $V_{\rm g}$) are filtered with Thermocoax cables and Cu powder filters, and the RF line has a low pass filter (Mini-Circuits VLFX-1350), a Cu powder filter and a $-40$\,dB attenuator at room temperature.

\subsection{Quasiparticle dominated excess current}
In this section we verify whether the excess current $\delta I = I - ef$ in the turnstile regime
of a superconducting hybrid single-electron transistor (SET), both of NISIN kind with the superconducting (S) island and of SINIS kind with the normal metal (N) island, is dominated by the quasiparticle contribution to the tunneling rates \cite{Saira2012, Maisi2014}.
Here `I' stands for insulating barrier.

From the theoretical side as shown in Supplementary Note 6 of \cite{Taupin2016} the excess current in both types of superconducting hybrid turnstiles takes the form of Eq.~(1) from the main text
\begin{equation}
\label{SM_eq:QP_current_general}
{\delta}I = C(A_{\rm g})\frac{\sqrt{2\pi\Delta_{\rm J}k_{\rm B}T_{\rm J}}}{eR_{T}}\exp\left[-\frac{\Delta_{\rm J}}{k_{\rm B}T_{\rm J}}\right] 
 \ ,
\end{equation}
where the overheating of S-parts is taken into account by the values of the superconducting gap $\Delta_{\rm J}$ and the electronic temperature $T_{\rm J}$ at the superconducting side close to the junction with the resistance $R_T$.
A certain form of the numerical coefficient $C(A_{\rm g})\sim 1$ considered, e.g., in \cite{Taupin2016} depends on the wave-form and the amplitude $A_{\rm g}$ of the RF gate voltage and considered as a fitting parameter of the model.

Other contributions to $\delta I$ 
such as Andreev tunneling, cotunneling, and Cooper-pair-electron cotunneling either do not depend on the drive amplitude, or their $A_{\rm g}$-dependence cannot be factorized as in Eq.~\eqref{SM_eq:QP_current_general}.

\begin{figure}[t]
\centering{
\includegraphics[width=0.49\columnwidth]{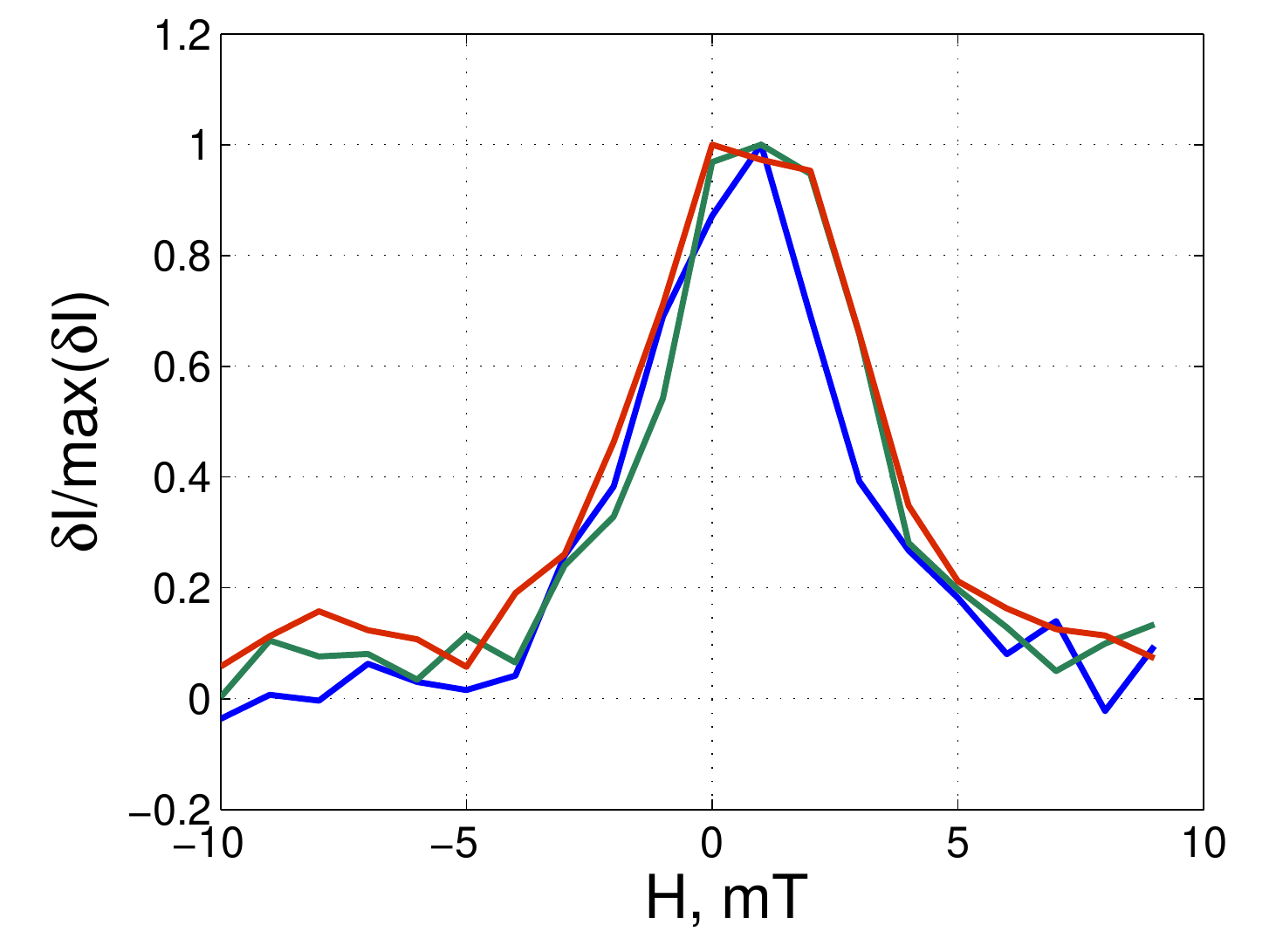}
\includegraphics[width=0.49\columnwidth]{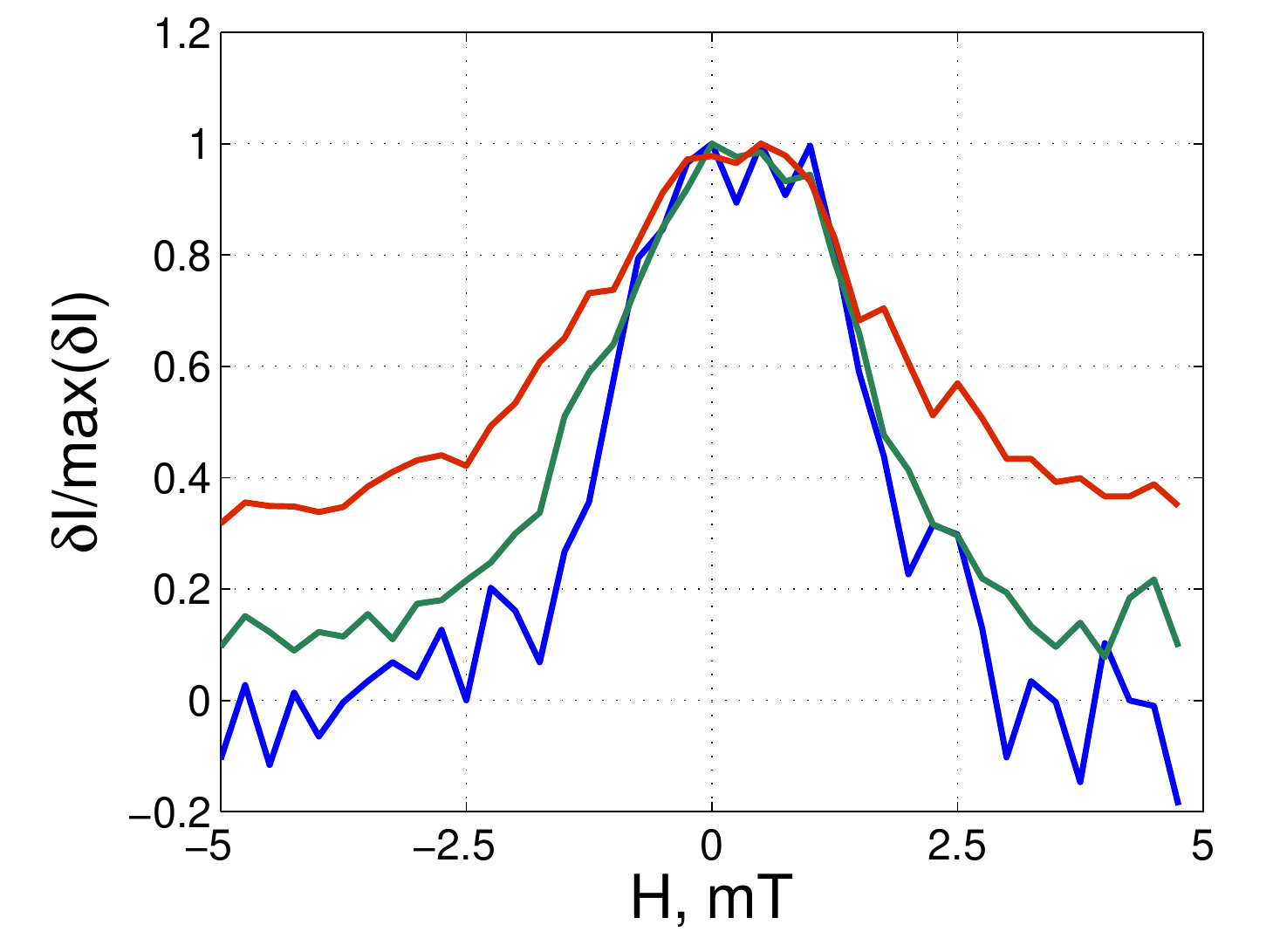}
\includegraphics[width=0.49\columnwidth]{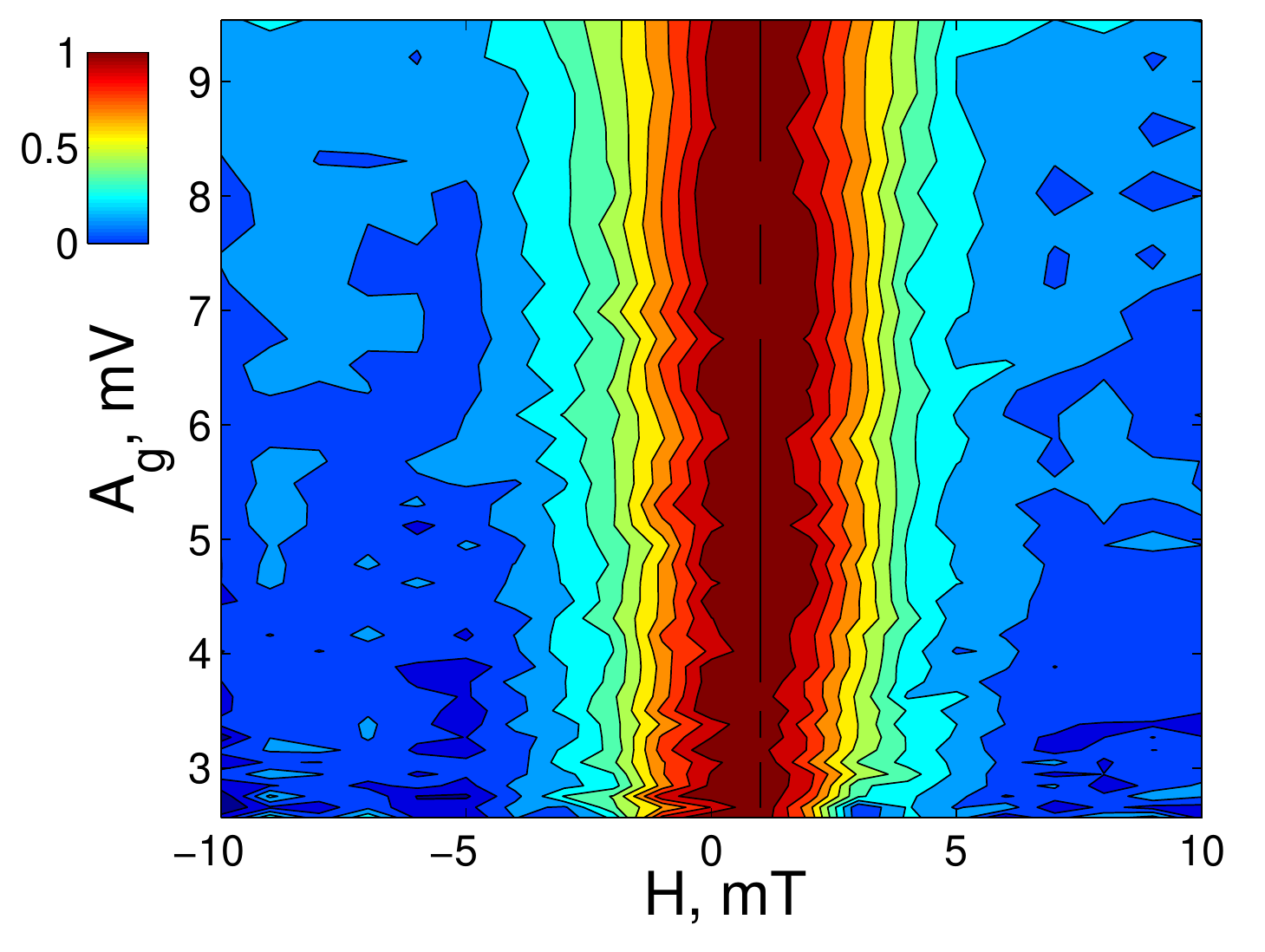}
\includegraphics[width=0.49\columnwidth]{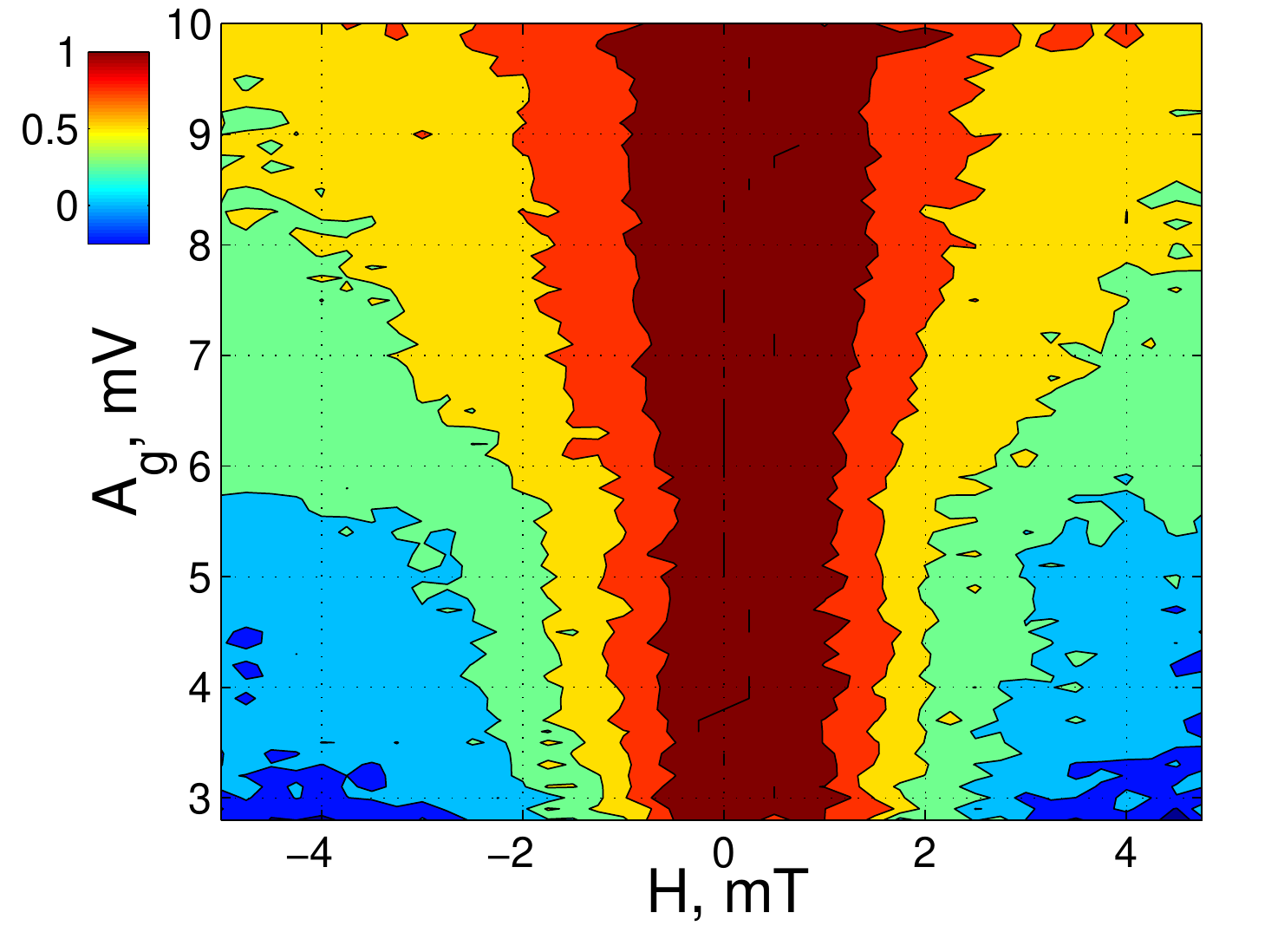}
}
\caption{The excess current $\delta I = I - e f$ in sample R (left panels) and sample L (right panels) normalized to its maximal value versus magnetic field (top panels) for different amplitude values $A_{\rm g} = 3$ (blue), $5$ (green), and 7\,mV (red).
Bottom panels show color plots at the turnstile plateau.
}\label{SM_Fig:dI_to_max(dI)_both Samples}
\end{figure}

Therefore to verify the dominant character of the quasiparticle contribution to the excess current
experimentally we normalized the magnetic field dependent data $\delta I(H,A_{\rm g})$ for different $A_{\rm g}$ values to its zero field value $\delta I(0,A_{\rm g})$, see Fig.~\ref{SM_Fig:dI_to_max(dI)_both Samples}.
The figure shows that in the sample R (left panels) the excess current scales with the drive amplitude $A_{\rm g}$ according to the theoretical factorizing formula \eqref{SM_eq:QP_current_general}, i.e., $\delta I (H,A_{\rm g}) = C(A_{\rm g}) \delta I_0(H)$.
However, in sample L (right panels) where the excess current is $\sim 10$ times smaller due to additional sample-holder shielding and wire filtering
(see Fig.~4(b) in the main text) other contributions play an important role at larger values of magnetic fields as the quasiparticle contribution is suppressed.
On one hand this additional shielding leads to smaller overheating effects (and therefore to the larger gap at the junction $\Delta_{\rm J}(0)$), but on the other hand, this causes the lower quality of the fitting in the sample L as the quasiparticle current \eqref{SM_eq:QP_current_general} is not the only contribution to the excess current in this case.

\subsection{Details of heat diffusion problem}
\begin{figure*}[h]
\centering{
\includegraphics[width=0.9\textwidth]{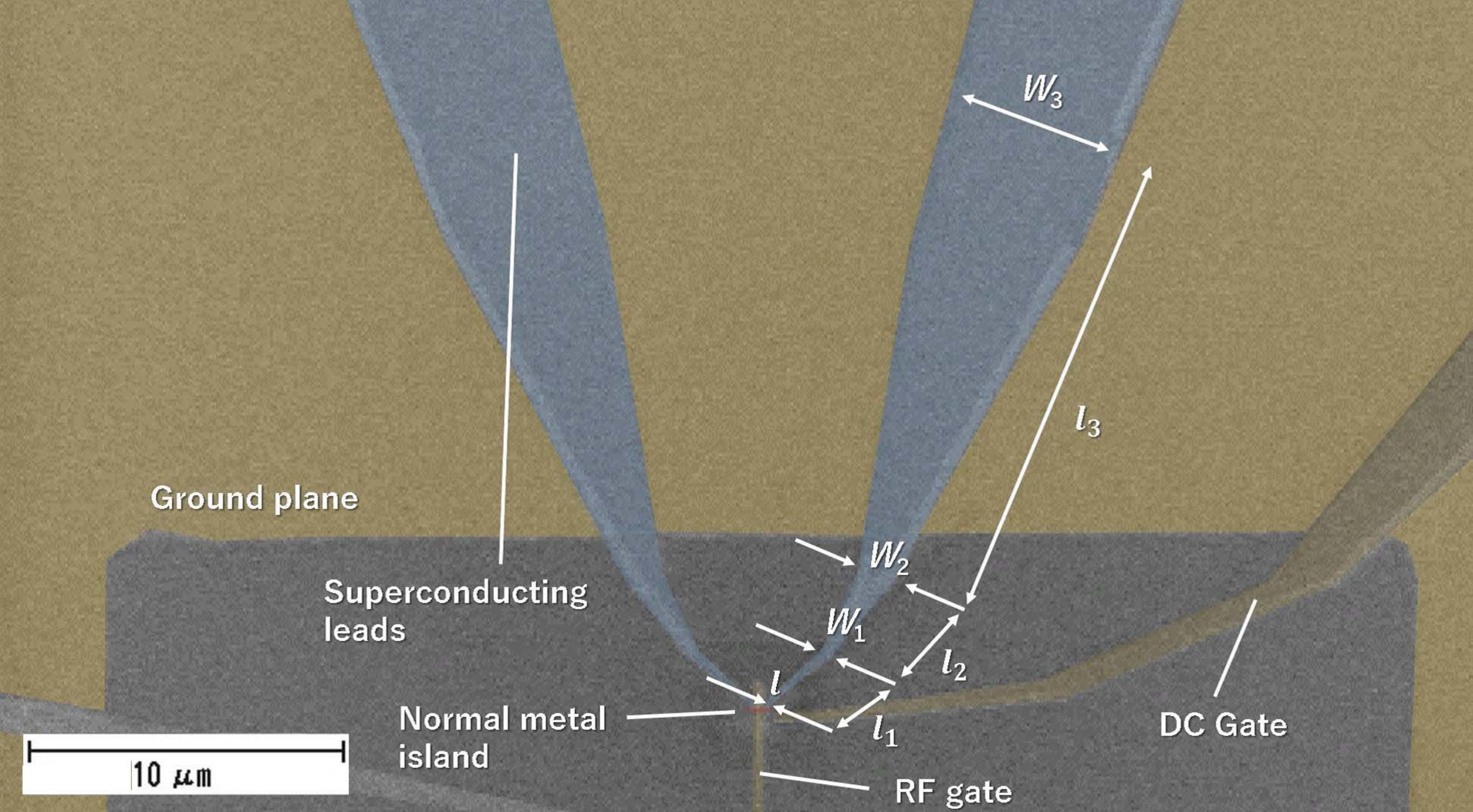}
}
\caption{SEM image of the SINIS SET shown in Fig.~1(a) of the main text at larger spatial scale.
As in the main text false color identifies superconducting Al leads (blue), the normal metal Cu island (red),
and the ground plane as well as the DC side gate and RF bottom gate electrodes (all yellow).
Widths $\ell$, $w_k$ and lengths $l_k$ mentioned in Eq.~\eqref{w(x)} are shown in the figure.
}\label{Fig:Layout}
\end{figure*}

Here we first verify the assumption (iii) of the main text and then give the details of the solution of the heat balance equation
within the assumptions (i-iv) mentioned in the main text and of the calculation of the normalized resistance $\mathcal{R}_H[w(x)]$ for a certain space profile of the lead width $w(x)$ of the considered samples, see Fig.~\ref{Fig:Layout}.

Starting with heat balance equation~(2) with boundary conditions~(3) from the main text we first consider the proximitized region $0<x<\ell$.
As this region is affected homogeneously by the inverse proximity effect we put $\Delta_H(x)=\Delta_{\rm J}$, i.e., $\kappa_{\rm J}[T(x)]\sim\frac{2 \Delta_{\rm J}^2}{e^2 \rho_n k_{\rm B}T(x)}e^{-\Delta_{\rm J}/k_{\rm B} T(x)}$ and neglect the electron-phonon relaxation $\dot{q}_{\rm e-ph}$ due to the smallness of the region $\ell\ll L_{T}$ compared to the electron-phonon relaxation length $L_T$.
This leads to the preservation of the heat flow in this region
\begin{gather}
\dot Q_\ell=\dot Q_{\rm S}\cdot A_\ell/A\simeq \dot Q_{\rm S}\cdot d_{\rm S}/\ell \ ,
\end{gather}
where $\dot Q_\ell$ and $A_\ell=\ell\cdot d_{\rm S}$ are the heat flow rate and the cross-sectional area of the S lead at $x=\ell$.
As a result the heat diffusion equation at $0<x<\ell$ takes the form
\begin{equation}\label{heat_diffusion_J}
\frac{\partial}{\partial x} \left[\kappa_{\rm J}\left(T(x)\right)\frac{\partial}{\partial x} T(x)\right] = 0 \ ,
\end{equation}
\begin{gather}
-\left.\kappa_{\rm J}\left(T(x)\right)\frac{\partial}{\partial x} T(x)\right|_{x=0}=\dot Q_{\rm S}/A \ .
\end{gather}
We consider the heat injection 
to be concentrated at one end $x=0$, but as we show below the concrete space distribution of
the heat injection does not matter.
Indeed, assuming the constant lead width in the region $x<\ell$ one can find that for $k_{\rm B} T(x)\ll \Delta_{\rm J}$
the quasiparticle number is linear decaying with coordinate
\begin{gather*}
\mathcal{N}_{qp}=\sqrt{\frac{2\pi k_{\rm B} T(x)}{\Delta_{\rm J}}} e^{-\frac{\Delta_{\rm J}}{k_{\rm B} T(x)}}\simeq\mathcal{N}_{qp}(0)-\frac{\dot Q_{\rm S} e^2 \rho_n x}{2 A k_{\rm B} T_{\rm J} \Delta_{\rm J}} \ .
\end{gather*}
In most of the cases the second term is negligible provided the $x<\ell\ll L_T$, therefore the assumption (iii) of the main text $\mathcal{N}_{qp}\simeq\mathcal{N}_{qp}(0)$ and $T(x<\ell)\simeq T_{\rm J}$ is valid.

To solve the heat balance equation (2) from the main text
in the rest of the lead $x>\ell$ we use continuity conditions for heat flow and temperature at the interface $x=\ell$
\begin{subequations}\label{T_bound_cond_w}
\begin{align}
\left.\kappa_{\rm J}\left(T\right)\frac{\partial}{\partial x} T(x)\right|_{x=\ell-0}&=\left.\kappa_{\rm S}(T,x)\frac{\partial}{\partial x} T(x)\right|_{x=\ell+0}
\ , \\
T_{\rm J}\equiv \left.T(x)\right|_{x=\ell-0}&=\left.T(x)\right|_{x=\ell+0} \ .
\end{align}
\end{subequations}

According to \cite{Knowles2012,Taupin2016} at rather small temperatures $T_0\ll T\lesssim 0.2 \Delta_H(x)/k_{\rm B} $
the density of the electron-phonon relaxation term can be approximated by the formula
$\dot{q}_{\rm e-ph} \sim \Sigma T^{5} e^{-\Delta_{H}(x)/k_{\rm B}T}$ neglecting the contributions proportional to $\propto e^{-2\Delta_{H}(x)/k_{\rm B}T}$ and $\propto e^{-\Delta_{H}(x)/k_{\rm B}T_0}$.
In this limit the main temperature dependence in both sides of the heat diffusion equation comes from the exponential $e^{-\Delta_{H}(x)/k_{\rm B}T}$ and this equation with logarithmic accuracy in terms of $x$-dependent quasiparticle number $\mathcal{N}_{qp}(x)$ takes form 
\begin{gather}\label{SM_eq:Nqp_diff_eq}
\frac{\partial^2}{\partial x^2}\mathcal{N}_{qp}(x) = L_T^{-2}\mathcal{N}_{qp}(x) \ ,
\end{gather}
with the unit diffusion coefficient and the weakly spatial dependent electron-phonon relaxation length $L_T^{-2}={e^2 \rho_n k_{\rm B}\Sigma T^4(x) }/{2\Delta_H(x)}$.
Separating the prefactor proportional to the heat injection source $\dot{Q}_{\rm S}$ one can write the solution of this equation 
\begin{gather}\label{n_qp_w}
\mathcal{N}_{qp}(\ell)\simeq\frac{\pi \dot Q_{\rm S} e^2 \rho_n}{\Delta_\ell \ell \sqrt{2 \pi k_{\rm B} T_{\rm J} \Delta_\ell}}\mathcal{R}_0\left[w(x)\right]
\end{gather}
in terms of the resistance $\mathcal{R}_H\left[w(x)\right]$ normalized to the sheet resistance depending on the lead width profile $w(x)$.

This normalized resistance $\mathcal{R}_H\left[w(x)\right]$ is the sum of terms $l_k/w_k$ and $\alpha^{-1}\ln(w_{k+1}/w_k)$ of one- and two-dimensional Green's functions of the Laplace equation
for the $k$th lead part of the length $l_k$ with constant width $w(x)=w_{k-1}$ and the linearly increasing one $w(x) = w_{k-1} + \delta x_k (w_k-w_{k-1})/l_k$, respectively. Here the coordinate $\delta x_k = x-x_k$ is shifted to be in the range $0<\delta x_k<l_k$ in the $k$th lead part and the opening angle is determined by $2\tan \alpha = (w_k-w_{k-1})/l_k$.

The lead geometry of the considered samples shown in Fig.~1(a) of the main text and in Fig.~\ref{Fig:Layout} leads to the following width profile
\begin{gather}\label{w(x)}
w(x)=\left\{
\begin{array}{ll}
\ell+\frac{w_1-\ell}{l_1}\delta x_0, & 0<\delta x_0<l_1\\
w_1+\frac{w_2-w_1}{l_2}\delta x_1, & 0<\delta x_1<l_2\\
w_2+\frac{w_3-w_2}{l_3}\delta x_2, & 0<\delta x_2<l_3\\
w_3, & 0<\delta x_3\\
\end{array}
\right.
\end{gather}
where $\delta x_k = x-\ell -\sum_{1}^k l_k$ and the junction region $0<x<\ell$ with $w(x)=\ell$ is followed by a set of linearly opening parts with corresponding lengths $l_k$ and width $w(x)$ changing from $w_{k-1}$ to $w_k$.

Taking the value estimates from the SEM micrographs $\ell\simeq 50$~nm, $w_1\simeq 275$~nm, $w_2\simeq 842$~nm, $w_3\simeq 8.4$\,$\mu$m, $l_1\simeq 2.85$\,$\mu$m, $l_2\simeq 2.4$\,$\mu$m, $l_3\simeq 28$\,$\mu$m and
using the estimate $L_T^2=\frac{R_T \ell^2 d_{\rm S}}{\rho_n}\sqrt{\frac{2k_{\rm B} T}{\pi\Delta_\ell}}\simeq (5$\,$\mu$m$)^2$ from \cite{Knowles2012} (in the presence of a normal shadow trap lying on top of the S lead far away from the junction with the same sheet NIS resistance) one can obtain at zero magnetic field
\begin{gather}\label{Est:G_0}
\mathcal{R}_0=\sum_{k=1}^{3}\frac{l_k}{w_k-w_{k-1}}\ln\left(\frac{w_k}{w_{k-1}}\right)+\frac{L_T}{w_3}\simeq 35 \ .
\end{gather}



To estimate $\mathcal{R}_H$ at finite magnetic field one should take into account the penetration of vortices into the lead at distances
$x>x_c$ where the S leads are already rather wide to let the first vortex to enter $w(x_c)\sim \sqrt{\Phi_0/\pi H}$ \cite{Stan_Phi0_w2}.
As the electron-phonon relaxation term in vortices is of the order of the one in the normal metal \cite{Taupin2016} they simply relaxes the temperature $T(x)\simeq T_0$ to its phonon bath value $T_0$.
Therefore at such distances $x>x_c$ one should truncate the summation in $\mathcal{R}_H$.

Taking this into account we estimate $\mathcal{R}_H$ at magnetic field $H\simeq 10$\,mT, when the critical width for the first vortex entry $w(x_c)\simeq \sqrt{\Phi_0/\pi H}\simeq 250$\,nm leads to the keeping of the first term in sum in Eq.~\eqref{Est:G_0} with $w_1$ substituted by $w(x_c)$ in the logarithm. It gives eventually
\begin{gather}\label{Est:G_H}
\mathcal{R}_H\approx\frac{l_1}{w_1-w_{0}}\ln\left(\frac{w(x_c)}{w_{0}}\right)\simeq 20 \ .
\end{gather}


\begin{thebibliography}{99}
\bibitem{Footnote_const_Nqp}
This is valid due to homogeneity of the heat inflow and the smallness of $\ell$ with respect to the electron-phonon relaxation length $L_{T}$.
\end{thebibliography}
\end{document}